\documentclass[
prl
,amsmath
,amssymb
,amsfonts
,twocolumn
,letterpaper
,10pt
,tightenlines
,nofootinbib
,nobalancelastpage
]{revtex4}
\usepackage{tikz}
\usepackage{subfigure}
\usepackage{qcircuit}
\usepackage{braket}
\usepackage{mathtools}
\usepackage{bm}
\usepackage{xcolor}
\usepackage{hyperref}

\graphicspath{figures/}

\allowdisplaybreaks[4]

\makeatletter

\newcommand{\ringplus}{\mathbin{\text{\@ringplus}}}

\newcommand{\@ringplus}{%
  \ooalign{\hidewidth\raise1.3ex\hbox{\tiny$\circ$}\hidewidth\cr$\m@th+$\cr}%
}

\newcommand{\ringminus}{\mathbin{\text{\@ringminus}}}

\newcommand{\@ringminus}{%
  \ooalign{\hidewidth\raise0.9ex\hbox{\tiny$\circ$}\hidewidth\cr$\m@th-$\cr}%
}
\makeatother
\newcommand{\tp}[0]{\mathrm{T}}

\DeclareFontFamily{U}{wncy}{}
\DeclareFontShape{U}{wncy}{m}{n}{<->wncyr10}{}
\DeclareSymbolFont{mcy}{U}{wncy}{m}{n}
\DeclareMathSymbol{\Sh}{\mathord}{mcy}{"58}

\newcommand{\negspace}{\!}
\newcommand{\lsub}[2]{{\protect\vphantom{#1}}_{#2} \negspace {#1}}

\newcommand{\lrsub}[3]{{\protect\vphantom{#1}}_{#2} \negspace {#1} \negspace {\protect\vphantom{#1}}_{#3}}

\newcommand{\brasub}[2]{\lsub {\bra{#1}} {#2}}

\newcommand{\pbra}[1]{\brasub{#1} p}

\newcommand{\avgg}[1]{\langle {#1} \rangle}

\newcommand{\op}[1]{\hat{#1}}

\newcommand{\opvec}[1]{\op{\vec{#1}}}

\newcommand{\id}[0]{I}

\newcommand{\mat}[1]{\bm{\mathrm{#1}}}

\renewcommand{\vec}[1]{\bm{\mathrm{#1}}}

\newcommand{\controlled}[1]{\op{\mathrm{C}}_{#1}}
\newcommand{\CZ}[0]{\controlled Z}

\newcommand{\blk}{\color{black}}

\usepackage{graphicx}
\graphicspath{{./figures/}}

\newcommand{\prlsection}[1]{\vspace{0.5ex}\textbf{#1}}

\newcommand{\subin}{{\text{in}}}
\newcommand{\subout}{{\text{out}}}
\newcommand{\subavg}{{\text{avg}}}

\newcommand{\sqzvar}{{\varepsilon}}
\newcommand{\asqzvarex}{{\kappa}}

\newcommand{\convsubsub}[2]{\mathrel{\lrsub {\,*\,} {#1} {#2}}}

\begin{document}

\title{Robust fault tolerance for continuous-variable cluster states with excess anti-squeezing}

\author{Blayney W. Walshe} 
\email{blayneyw@gmail.com}
\author{Lucas J. Mensen}
\author{Ben Q. Baragiola}
\author{Nicolas C. Menicucci}
\affiliation{Centre for Quantum Computation and Communication Technology, School of Science, RMIT University, Melbourne, VIC 3000, Australia}

\begin{abstract}
The immense scalability of continuous-variable cluster states motivates their study as a platform for quantum computing, with fault tolerance possible given sufficient squeezing and appropriately encoded qubits  [Menicucci, PRL~\textbf{112}, 120504 (2014)]. Here, we expand the scope of that result by showing that additional anti-squeezing has no effect on the fault-tolerance threshold, removing the purity requirement for experimental continuous-variable cluster-state quantum computing. We emphasize that the appropriate experimental target for fault-tolerant applications is to directly measure 15--17~dB of squeezing in the cluster state rather than the more conservative upper bound of 20.5~dB.
\end{abstract}
\maketitle

\prlsection{Introduction.}---%
Measurement-based quantum computation (MBQC) employs highly entangled resource states known as cluster states~\cite{Briegel2001} as a substrate for a quantum computation~(QC). A specific computation is carved from the cluster state using only adaptive single-qubit measurements~\cite{Raussendorf2001}. MBQC eliminates the need for coherent, multi-qubit interactions during the computation, which provides an advantage over circuit-model methods~\cite{Nielsen2010}.

Continuous-variable~(CV) MBQC extends this scheme by utilising CV resources, rather than qubits, to build the initial cluster state~\cite{Nick2006}. This provides a distinct advantage in scalability over optical-qubit-based MBQC schemes since CV cluster states can be made deterministically on an immense scale~\cite{Menicucci2008,Menicucci2011a}. Other MBQC schemes have been able to achieve qubit cluster states of only 6 qubits utilising two photons \cite{Ceccarelli2009}, compared with 60 frequency modes~\cite{Chen:2014jx} or $10^4$--$10^6$ temporal modes for a CV cluster state~\cite{Yokoyama2013, Yoshikawa2016}, albeit with a 1D topology in both cases. Accessible proposals exist for making computationally universal (i.e., 2D) CV cluster states on a similar scale~\cite{Menicucci2011a,Wang:2014im,Alexander2016,Alexander2018}.

Still, \blk finite squeezing (required by finite energy) deposits noise that accumulates throughout a computation~\cite{Alexander2014,Gu2009}. Appropriately encoded qubits~\cite{GKP}---available on demand to be coupled into the CV cluster state at will---can survive this noise with regular rounds of quantum error correction. As long as the squeezing---in both the CV cluster state and in the encoded qubits---is high enough, fault-tolerant quantum computation is possible%
~\cite{Nick2014}. 

The main idea behind that result is to convert the additive Gaussian noise due to finite squeezing~\cite{Gu2009} to logical-Pauli noise at the encoded-qubit level after every logical-Clifford gate. The quantum-error-correction scheme proposed by Gottesman, Kitaev, and Preskill~(GKP)~\cite{GKP} enables this. 
With the logical-Clifford gates regularly spaced into a grid and supplemented by distillation of magic states~\cite{Bravyi:2005dx},
the problem is effectively mapped to a circuit-model computation using noisy gates~\cite{Nick2014}. This is a well-studied problem (see Ref.~\cite{Gottesman:2009ug} for a review). The acceptable \emph{threshold} for gate errors depends on the chosen qubit-level quantum error-correcting code employed to get rid of this residual error.

\begin{figure}[t!]
\includegraphics[width=\columnwidth]{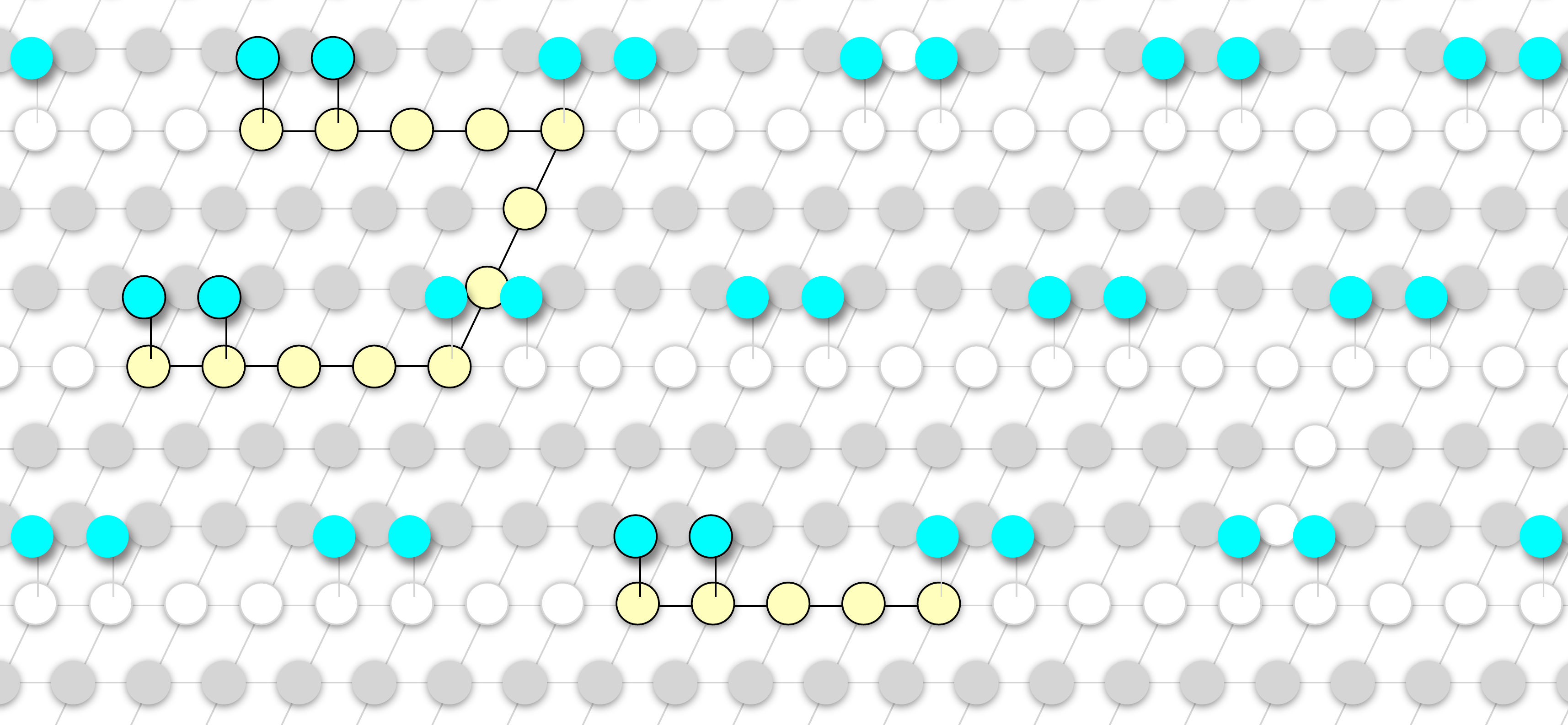}
\caption{\label{fig:flowerbed}
CV resource state for fault-tolerant quantum computation. In the original proposal, Ref.~\cite{Nick2014}, the base layer consists of nodes that are momentum-squeezed vacuum states, with the lines connecting them representing $\CZ[1] = e^{i \op q \otimes \op q}$ gates. Here, we consider each node to be an impure momentum-squeezed thermal state, which has additional noise in the position quadrature. The cyan nodes are GKP-encoded~$\ket {0_L}$ states~\cite{GKP}, attached via $\CZ[1]$ gates at regular intervals like flowers in a flowerbed. These are used for error correction, while the grey and white nodes implement the computation (see text). The yellow highlighted nodes represent two-mode (left) and one-mode (right) error-corrected gates, described in more detail in Fig.~\ref{fig:gates}.
}
\end{figure}

Figure ~\ref{fig:flowerbed} depicts a CV resource state for universal computation. We describe this as a ``flowerbed'' comprising two essential parts: (1)~a large, canonical~\cite{Menicucci2011} CV cluster state with a square-lattice graph~\cite{Gu2009}; and (2)~GKP-encoded $\ket{0_L}$ ancillae attached at regular intervals to the cluster-state base by $\CZ[1] = e^{i \op q \otimes \op q}$ gates. Performing $\op q$ measurements on unwanted nodes (grey) ``carves out'' the structure of a desired CV quantum circuit. The remaining nodes are measured in $\op p$ or $\op p + \op q$ (known as a \emph{shear measurement}~\cite{Nick2014}) to enact one- and two- qubit logical Clifford gates. Universality is achieved by distillation of magic states~\cite{Bravyi:2005dx}, which was previously thought to require an additional non-Gaussian resource~\cite{GKP,Nick2014}. It was recently shown, however, that distillable GKP magic states can be produced using heterodyne detection---a Gaussian measurement---on a GKP-encoded Bell pair~\cite{Baragiola2019}.

\begin{figure}[t!]
\includegraphics[width=0.9\columnwidth]{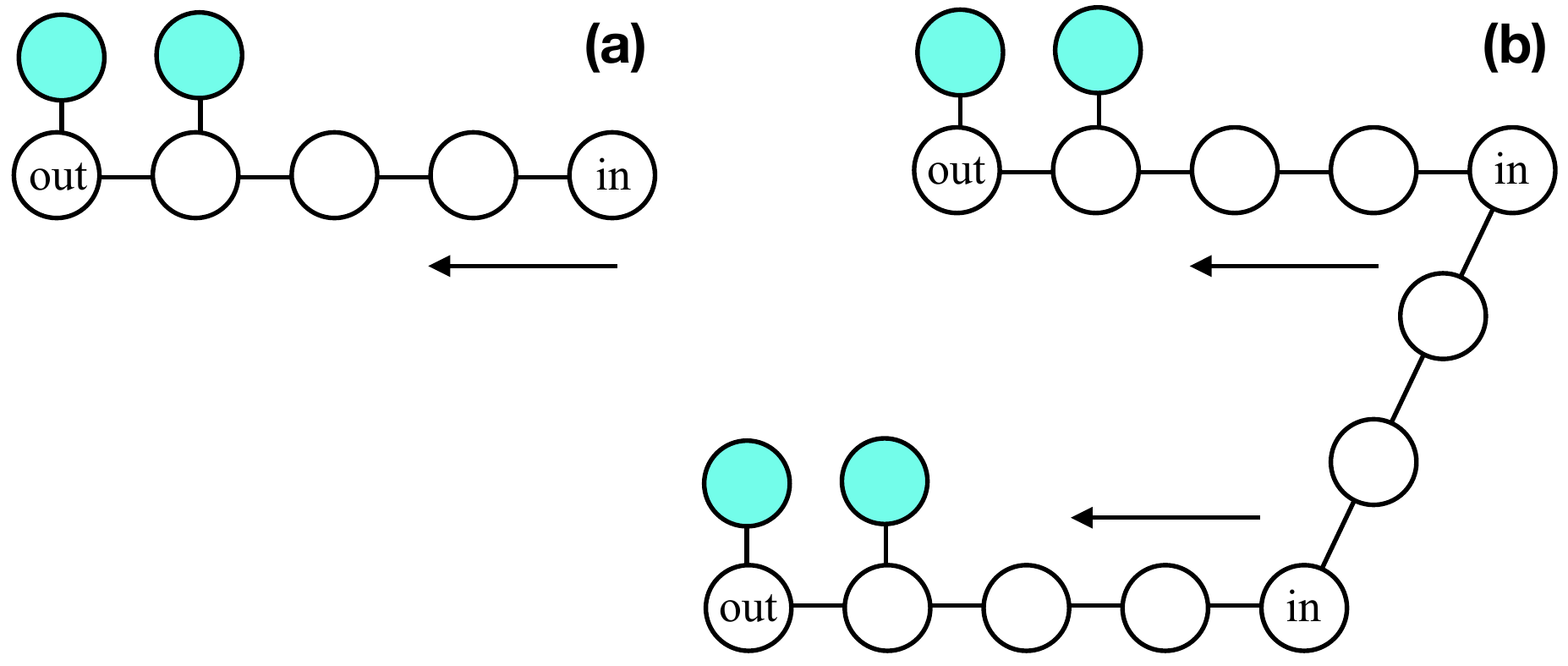}
\caption{\label{fig:gates} The subgraphs sufficient to enact any (a)~single-qubit and (b)~two-qubit logical-Clifford gate, followed by GKP error correction~\cite{Nick2014}. These have been cut from the flowerbed using $\op q$ measurements on the adjacent nodes. The cyan nodes are GKP~$\ket {0_L}$ for error correction~\cite{GKP}, the blank nodes are squeezed thermal states [Eq.~\eqref{eq:therm}], \emph{in} is the output node of the previous gate, and the arrows indicate the direction of the measurement sequence.}
\end{figure}

A single example of both the one- and two-mode gates is shown in yellow and again independently in Fig.~\ref{fig:gates}. The output of one such gate becomes the input of the next. This way, the white nodes in Fig.~\ref{fig:flowerbed} can be seen as a series of one- or two-mode gates acting in sequence. With this procedure, the one-mode gate is sufficient to enact any single-mode Gaussian unitary gate by a series of shear measurements followed by error correction.

The gate in Fig. \ref{fig:gates}(b) is required to enact two-mode quantum gates such as the $\CZ$ gate. This gate functions by connecting the two input nodes vertically with two nodes to be measured in~$\op p$. Measuring these connecting nodes implements a $\CZ[-1] = e^{-i \op q \otimes \op q}$ gate on the input states, after which they pass through two one-mode identity gates (which are included only to keep the calculation on a regular lattice~\cite{Nick2014}).

One of us~\cite{Nick2014} has demonstrated that for a pure cluster state (i.e., one created from squeezed vacuum states), there exists a finite squeezing threshold of no higher than 20.5 dB that enables fault-tolerance to be achieved. Here we generalize that result to the case where the CV cluster-state (base of the flowerbed) is built from squeezed \emph{thermal} states instead of squeezed vacuum states. This introduces additional anti-squeezing, which will turn out---surprisingly---not to affect the threshold calculations at all.

\prlsection{Definitions.}---%
We work with quadrature operators ${\op q = \frac {1} {\sqrt 2} (\op a + \op a^\dag)}$ and ${\op p = \frac {-i} {\sqrt 2} (\op a - \op a^\dag)}$ satisfying ${[\op q, \op p] = i}$, with ${\hbar = 1}$. The vacuum variance is ${\avgg{\op q^2} = \avgg{\op p^2} = \frac 1 2}$. We denote column vectors of position and momentum operators as~$\opvec q$ and $\opvec p$, respectively, and we collect both into the column vector~$\opvec x \coloneqq (\opvec q^\tp, \opvec p^\tp)^\tp$.

A \emph{squeezed vacuum state} with \emph{squeezing factor}~${s > 1}$ is a Gaussian state with 0~mean, variance~${\sqzvar_0 \coloneqq \frac 1 2 s^{-2}}$ along the squeezed quadrature, and variance~$\asqzvarex_0 \coloneqq \frac 1 2 s^2$ along the anti-squeezed quadrature. The corresponding \emph{squeezing parameter}~${r = \frac 1 2 \ln s}$ so that ${s = e^{2r}}$. Note that a measured variance~$\sigma^2$ corresponds to $10 \log_{10} (2 \sigma^2)$~dB, with negative corresponding to squeezed and positive to anti-squeezed.

\begin{figure}[t!]
\includegraphics[width=0.4\textwidth]{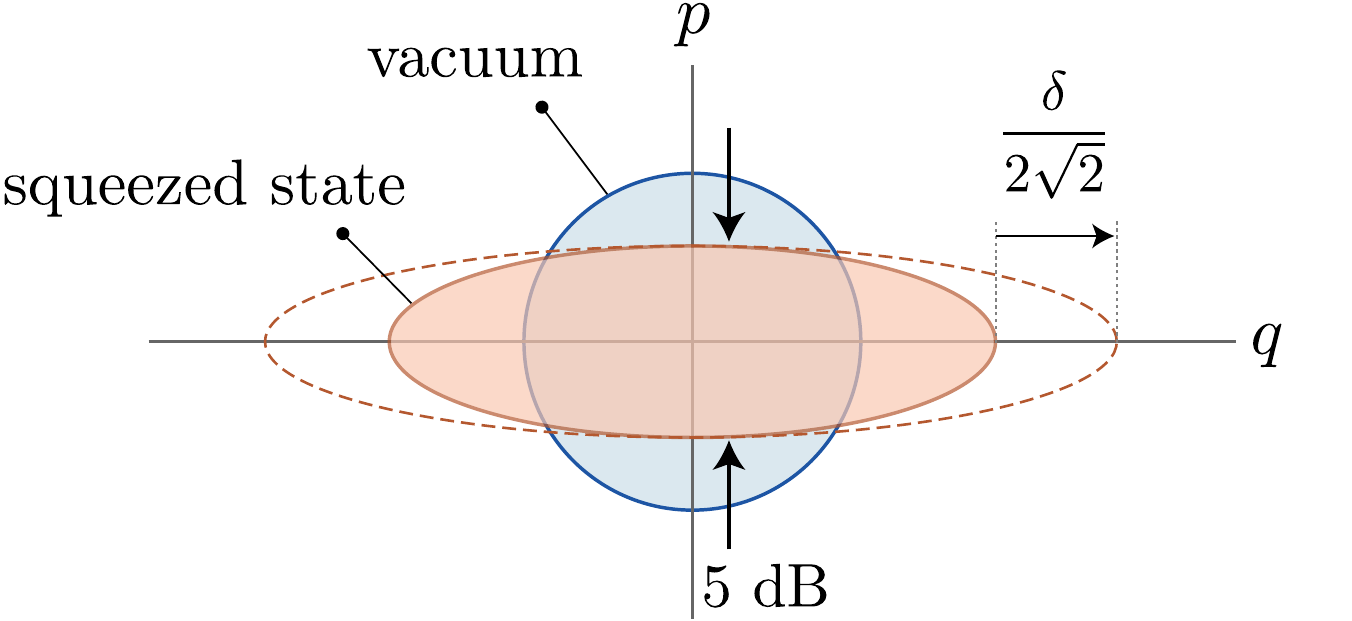}
\caption{
Phase-space comparison of the vacuum ($s=1$) and a 5-dB $p$-squeezed vacuum state ($s=1.78$). Additional anti-squeezing in the $q$ quadrature models a squeezed thermal state (dashed). Shown are 1-$\sigma$ error ellipses for the thermal-state Wigner functions, Eq.~\eqref{eq:therm}.}
\label{fig:image}
\end{figure}

A \emph{squeezed thermal state} is defined here in terms of its measured variances rather than in terms of a squeezing parameter and a temperature. Using this convention, we designate the variance~$\sqzvar$ along the squeezed quadrature to match the squeezed-vacuum case---i.e.,~$\sqzvar = \sqzvar_0 = \frac 1 2 s^{-2}$, while the variance along the anti-squeezed quadrature is larger than in that case: ${\asqzvarex \coloneqq {\frac 1 2 (s^2 + \delta^2)} = \asqzvarex_0 + \frac 1 2 \delta^2}$. The additional variance in the anti-squeezed quadrature, $\frac 1 2 \delta^2$, is called the \emph{additional anti-squeezing}. The Wigner function for this state is
\begin{align}
\label{eq:therm}
	W_{\asqzvarex,\sqzvar}(q,p)
&
\coloneqq
	G_{\asqzvarex}(q) G_{\sqzvar}(p) 
=
	G_{\frac{1}{2}(s^2+\delta^2)}(q) G_{\frac{1}{2}s^{-2}}(p) 
	,
\end{align}
where $G_{\sigma^2}(x)$ is a normalised Gaussian with zero mean and variance~$\sigma^2$. Note that $\sqrt{\sqzvar \asqzvarex} \geq \frac 1 2$, with equality if the state is pure (squeezed vacuum), and $\sqzvar = \asqzvarex = \frac 1 2$ in the case of the vacuum. These states are shown in Fig.~\ref{fig:image}.

\prlsection{Wigner representation of cluster-state computation.}---%
As quantum information propagates from node to node through a CV cluster state, operations are performed by projective measurements on each node. Due to finite squeezing, noise is introduced at each of these steps, which appears as convolution of one quadrature of the input-state Wigner function when the cluster state is pure (made from squeezed vacuum)~\cite{Alexander2014,Gu2009}. 

By tracking how this noise accumulates throughout the computation, we can determine how the additional anti-squeezing might affect the qubit-level error rate. This is achieved, as discussed throughout the supplementary material in Ref.~\cite{Nick2014},  by evolving the input Wigner function through the appropriate quantum gates according to $W(\vec x) \xrightarrow{\text{$\op G$}} W' \big[\mat S^{-1}_{\op G}(\vec x-\vec c) \big]$, where $\vec x \coloneqq (\vec q^\tp, \vec p^\tp)^\tp$ is the vector of all phase-space coordinates, and where $\mat S_{\op G}$ and~$\vec c$ are found via the Heisenberg action of the Gaussian unitary~$\op G$ on the vector of quadrature operators~$\opvec x$---i.e., $\hat{G}^{\dag} \opvec x \hat{G} = \mat S_{\hat{G}} \opvec x +  \vec c$ \cite{Alexander2014}. Quadrature measurements replace the measured variable with its outcome and integrate over the conjugate variable---e.g., measuring $\op p_1$ with outcome~$s$ maps $W(\vec x) \to \tilde{W}_{\text{out}}(\vec x_{\geq 2}) \coloneqq \int dq_1\, W(\vec x)\rvert_{p_1 = s},$ where $\vec x_{\geq 2}$ is $\vec x$ for the unmeasured modes, and the tilde indicates that the Wigner function is unnormalised.

\prlsection{Results.}---%
We examine four occasions where additional anti-squeezing might affect the cluster-state output: (1)~using $\op q$ measurements to delete a node, (2)~a one-mode gate, (3)~a two-mode gate, and (4)~magic state preparation.

(1)~\emph{Deletion via $\op q$ measurements.}---%
Since each node of the flowerbed is attached to its neighbors by a $\CZ[1]$ gate, we only need to use the fact that a $\op q$ measurement after this gate just induces an outcome-dependent momentum shift on the other mode: $(\brasub s {q_1} \otimes \op \id) e^{i \op q_1 \op q_2} = e^{i s \op q_2} = \op Z_2(s)$. Since~$s$ is known, we can correct it on each neighboring node with~$\op Z(-s)$, and the result is the same as if the deleted node had never been attached in the first place. The input state makes no difference to this analysis.

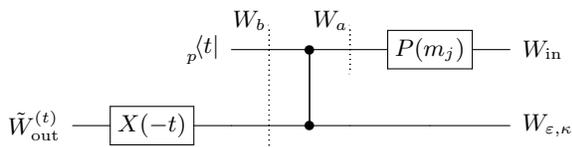
\begin{figure}[t!]
\begin{align*}
	\Qcircuit @C=0.5cm @R=.5cm {
		& & & \lstick{\pbra{t}}   &  \ustick{ ^{\displaystyle W_b}\;\;\quad}
		\qw & \ctrl{1}    &  \ustick{ ^{\displaystyle W_a}\;\;\quad}\qw & \gate{P(m_j)} &  \rstick{ W_\subin} \qw \\
		& \lstick{ \tilde W_\subout^{(t)}}  & \gate{X(-t)} &\qw & \qw & \ctrl{-1}  & \qw & \qw & \rstick{ W_{\sqzvar,\asqzvarex}} \qw 
		\POS"1,5"."2,5"!C*+<0em,2em>\frm{.}
		\POS"1,7"."1,7"!C*+<0em,2em>\frm{.}
		}
\end{align*}
\caption{The quantum circuit (read right to left) describing one measurement and outcome-dependent displacement as part of the single-mode gate shown in Fig.~\ref{fig:gates}(a). $W_\subin$ is the Wigner function for an arbitrary quantum state, and $W_{\sqzvar,\asqzvarex}$ is that for the squeezed thermal state [Eq.~\eqref{eq:therm}] at the next node in the cluster. $W_a$ is the input state after the shear gate, and $W_b$ is the two-mode Wigner function after the $\CZ[1]$ gate. The unnormalized, outcome-dependent output state is represented by~$\tilde W_\subout^{(t)}$.
}
\label{singlecircuit}
\end{figure}

(2)~\emph{One-mode gate.}---%
Recall Fig.~\ref{fig:gates}(a). A one-mode gate requires several measurements of either $\op p$ or $\op p + \op q$. For mode~$j$, we write each measurement as $\op p_j + m_j \op q_j$ and note that ---in principle---we could have implemented either measurement, $m_j \in \{0,1\}$, by performing an $m_j$-dependent shear gate~$\op P(m_j) = e^{i m_j \op q^2/2}$ before measuring~$\op p$~\cite{Nick2014}, and the output would be the same as what we actually do---i.e., measure $\op p_j + m_j \op q_j$. This alternate picture of placing an $m_j$-dependent shear gate before a fixed measurement will assist with the analysis, and the corresponding circuit for this is shown in Fig.~\ref{singlecircuit}.

The input to the circuit is~$W_\subin$. For later use, we define a new Wigner function~$W_a$ to represent the input state after having passed through the shear gate~$\op P(m_j)$ in Fig.~\ref{singlecircuit}. This is just
\begin{align}
	W_a(q_1,p_1) = W_\subin(q_1, p_1 - m_j q_1)
	\,.
\end{align}
The $\CZ[1]$ gate entangles $W_a$ and $W_{\sqzvar,\asqzvarex}$ to give
\begin{align}
\label{eq:singlemode1}
	W_b(\vec x)
&
=
	W_a(q_1,p_1 - q_2)
	G_\asqzvarex(q_2)G_\sqzvar(p_2-q_1)
	,
\end{align}
and the outcome-dependent final state after the measurement and displacement is~\cite{Alexander2014,Gu2009}
\begin{align}\label{eq:singleout}
	\tilde{W}^{(t)}_\subout(q_2,p_2) &= G_\asqzvarex(q_2+t) [G_\sqzvar \convsubsub {} 1 W_a](p_2, -q_2),
\end{align}
where $\convsubsub {} 1$ indicates convolution of the univariate function on the left with the 1st variable of the function on the right~\cite{Nick2014}---e.g.,
\begin{align}\label{eq:conv}
[f \convsubsub {} 1 g] (x,y) \coloneqq \int dw \, f(w)g(x-w,y)
\end{align}
and similarly for the other $\convsubsub {} n$ using the $n$th argument of~$g$.

In Eq. \eqref{eq:singleout} the input Wigner function is being convolved by a narrow Gaussian~$G_\sqzvar$ in its first argument, which can be interpreted as a slight blurring in that quadrature, and then it is multiplied by a wide Gaussian envelope~$G_\asqzvarex(q_2+t)$. This envelope holds all dependence on the additional anti-squeezing through its variance~$\asqzvarex$. Its mean depends on the outcome~$t$ after the final displacement.

\begin{figure}[t!]
\begin{align*}
	\raisebox{-4.5em}{ $\tilde{W}_\subout^{(r,t)}$\;\;}
	\Qcircuit @C=1.4em @!R=0.5em {
		 & \gate{Z(-t)}    & \qw    & \ustick{ ^{\displaystyle W_a}\;\;\quad}\qw    & \qw     & \ctrl{1} & & & & \ustick{\quad\quad _{(1)}} &
		 \qw[-5] & \\
		 & & \lstick{\pbra{r}}  & \qw    & \ctrl{1} & \ctrl{-1}  & \ustick{\quad _{(2)}} & \rstick{ W_{\sqzvar,\asqzvarex} } \qw[-2] \\
		& & \lstick{\pbra{t}}  & \qw    & \ctrl{-1} & \ctrl{1}  & \ustick{\quad _{(3)}} & \rstick{ W_{\sqzvar,\asqzvarex} } \qw[-2] \\
		 & \gate{Z(-r)}   & \qw   & \qw        & \qw     & \ctrl{-1} & & & & \ustick{\quad\quad _{(4)}} &
		 \qw[-5] & \ 
		\gategroup 1 1 4 1 {1.5em} {\{}
		\gategroup 1 {11} 4 1 {1.5em} {\}}
		\POS"1,4"."4,4"!C*+<0em,2em>\frm{.}
		}	
	\raisebox{-4.5em}{ $W_\subin$\;\;}	
\end{align*}
\caption{The quantum circuit (read right to left) representing the action of $\op p$ measurements on the two nodes connecting the input states in Fig.~\ref{fig:gates}(b). The measurement-dependent correction step is included. $W_\subin$ is the Wigner function for an arbitrary two-mode input state, and $W_{\sqzvar,\asqzvarex}$ represents the squeezed thermal state [Eq.~\eqref{eq:therm}] at each of the two connecting nodes in the cluster. $W_a$ is the total Wigner function after the $\CZ[1]$ gates. The unnormalized, outcome-dependent output state is represented by~$\tilde W_\subout^{(r,t)}$.} 
\label{twocircuit}
\end{figure}
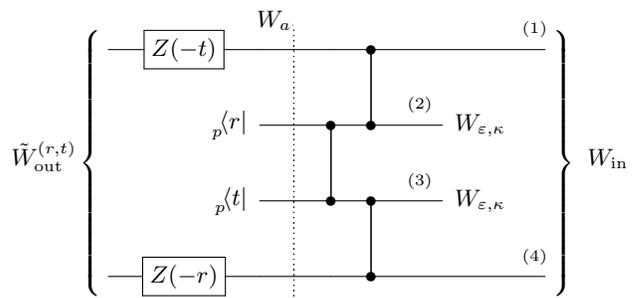

To model the typical effect of the channel~\cite{Gu2009}, we average the normalized state over the measurement outcome~$t$, which is equivalent to integrating the unnormalised Wigner function~$\tilde W_\subout^{(t)}$ over~$t$. This eliminates~$G_\asqzvarex$, giving the normalized Wigner function
\begin{equation}
	W^\subavg_\subout(q_2,p_2)
=
	[G_\sqzvar \convsubsub {} 1 W_a] (p_2,-q_2)
	.
\end{equation}
All dependence on the additional anti-squeezing is gone. In fact, this is exactly the same average effect as one gets with squeezed vacuum instead~\cite{Nick2014,Gu2009}. Only the variance~$\sqzvar$ of the squeezed quadrature affects the output (through convolution with~$G_\sqzvar$).

(3)~\emph{Two-mode gate.}---%
To determine the effects of the additional anti-squeezing on the two-mode gate in Fig.~\ref{fig:gates}(b) we need only consider the effect of $\op p$ measurements on the two nodes connecting the input states. This is because any subsequent measurements along the top or bottom wire are already covered by the result given above for the one-mode gate. The circuit for these two measurements is given in Fig.~\ref{twocircuit} 

The input state is a general two-mode state~$W_\subin(q_1, q_4, p_1, p_4)$ over modes 1 and~4. After the $\CZ[1]$ gates, the 4-mode state is
\begin{align}
	W_a(\vec x)
&
=
	W_\subin(q_1, q_4, p_1 - q_2, p_4 - q_3)
	G_\asqzvarex(q_2)
	G_\asqzvarex(q_3)
\nonumber
\\
&
\quad\times
	G_\sqzvar(p_2-q_1-q_3)
	G_\sqzvar(p_3-q_4-q_2)
	.
\end{align}
After the measurements and a change of integration variables to absorb the outcomes~$r$ and~$t$, the output state becomes
\begin{align}
&
	\tilde{W}_\subout^{(r,t)}
	(q_1, q_4, p_1, p_4)
\nonumber
\\
&
\quad =
	\int du\, dv\, W_\subin(q_1, q_4, p_1 + u, p_4 + v)
\nonumber
\\
&
\quad\quad\times
	G_\asqzvarex(t - u)
	G_\asqzvarex(r - v)
	G_\sqzvar(v - q_1)
	G_\sqzvar(u - q_4)
	.
\end{align}
Analogous to the single-mode calculation above, this involves two convolutions and two Gaussian envelopes. The latter hold the entire effect of the additional anti-squeezing, and once again they integrate to~1 after averaging over measurement outcomes, leaving
\begin{align}
&
	W_\subout^\subavg
	(q_1, q_4, p_1, p_4)
\nonumber
\\
&
=
	[G_\sqzvar \convsubsub {} 4 (G_\sqzvar \convsubsub {} 3 W_\subin)]
	(q_1, q_4, p_1 + q_4, p_4 + p_1)
	.
\end{align}
The result is just~$W_\subin$ acted upon by~$\CZ[-1]$ and then blurred by~$G_\sqzvar$ in the momentum quadrature of modes 1 and 4. This is the same as what happens with squeezed vacuum~\cite{Nick2014}, showing again that the final result is independent of the additional anti-squeezing.

(4)~\emph{Magic state preparation.}--- The supplementary material of \cite{Nick2014} discusses the preparation of magic states using photon counting on a GKP-encoded Bell state. Recent work has shown that heterodyne detection (which is Gaussian) can be used instead of photon counting (non-Gaussian) to produce the same results, enabling fault-tolerant, universal QC \cite{Baragiola2019} given a CV cluster state and a supply of encoded $\ket{0_L}$. The GKP-encoded Bell state is prepared using encoded Clifford gates, and then heterodyne detection is performed on one mode. If necessary, distillation of higher-quality magic states then proceeds using further Clifford circuits. We have shown in (3) that the additional anti-squeezing has no effect on the output of the two-mode gate and so does not affect magic-state preparation either.

In summary, we have taken fault-tolerant continuous-variable MBQC and demonstrated that this fault tolerance is maintained even in the presence of additional anti-squeezing. This result is significant in determining what error-correction steps are required for practical implementations of this QC scheme.

\prlsection{Discussion.}---%
The goal of Ref.~\cite{Nick2014} was to prove that a finite squeezing threshold exists for CV MBQC, not to optimize the particular threshold value. In fact, taking a conservative approach to the required squeezing ensured that this goal was achieved even if many of the implementation details (such as particular code to be used) were left unspecified.
The conservative squeezing threshold explicitly quoted, 20.5~dB, corresponds to a qubit-gate error rate of $10^{-6}$. This satisfies the most stringent threshold required by any known quantum-error-correcting code~\cite{Aharonov:1997:FQC:258533.258579}, making the result very general.

In fact, 20.5~dB is an \emph{upper bound} on the actual squeezing threshold for any particular code~\cite{Nick2014}, and lower squeezing is likely to be sufficient for particular applications. For instance, specifying the 23-qubit Golay code or 7-qubit Steane code allows for gate error rates of $\sim$$10^{-3}$~\cite{Steane2003}. This translates to~$\sim$17.4~dB of squeezing at the physical level using the construction in Ref.~\cite{Nick2014}. Using the $C_4/C_6$ code, which has a threshold of $\sim$1\% under conservative assumptions~\cite{Knill2005}, the construction of Ref.~\cite{Nick2014} gives a squeezing threshold of~$\sim$15.6~dB.

Using these levels of squeezing in a fault-tolerant QC will require more careful design of the algorithm to be implemented since some proofs of these qubit-based thresholds assumed aspects of circuit design that are prohibitive in a 2D cluster-state architecture (e.g., Ref.~\cite{Steane2003} assumes any-to-any 2-qubit gates), so additional theoretical work may be required to prove that a particular squeezing level will suffice when employing a particular code for a particular application.

Nevertheless, the present work shows a critical simplification: the anti-squeezing levels have no effect on any squeezing threshold calculated using the methodology of Ref.~\cite{Nick2014}. This result, while surprising on the surface, agrees with prior work on CV teleportation, which showed that additional anti-squeezing has no affect on teleportation fidelity~\cite{Takeda2013}. It would have an effect, however, on any attempt to use gain tuning to alter the noise model from additive Gaussian noise to, say, pure loss~\cite{Polkinghorne1999,Ralph1999,Takeda2013,Alexander2017} since that simplification requires a pure resource state~\cite{Takeda2013}. 

For completeness, we note that if one were to forego the scalability advantage of large-scale CV cluster states and instead make a CV resource state entirely out of GKP states, a lower threshold of~10~dB is possible---albeit daunting~\cite{Fukui2018}. To obtain this, Fukui \emph{et al.} employed analogue quantum error correction~\cite{Fukui2017}, postselection, and a 3D cluster-state construction using GKP input states~\cite{GKP}. This is an impressive threshold, but it does not immediately apply to methods based on large-scale CV cluster states, which have been experimentally demonstrated~\cite{Yokoyama2013, Yoshikawa2016, Chen:2014jx}.

For that purpose, the fault-tolerance target should be to directly measure 15--17~dB of squeezing in a CV cluster state. This squeezing can be converted to an effective qubit-level gate error rate (for use with GKP states~\cite{GKP} at similar squeezing levels) using the methodology of Ref.~\cite{Nick2014}. The present work shows that the level of anti-squeezing in the cluster state is irrelevant, thereby greatly expanding the scope of that result to include \emph{physical} (i.e., impure) CV cluster states.

\prlsection{Acknowledgments.}---%
We thank Rafael Alexander and Giacomo Pantaleoni for discussions. This work is supported by the Australian Research Council Centre of Excellence for Quantum Computation and Communication Technology (Project No.\ CE170100012).

\bibliographystyle{bibstyleNCM_papers_notitle}
\bibliography{references}

\end{document}